\documentstyle[floats,preprint,aps,epsf]{revtex}
\tighten

\begin{document}
\draft


\title{Calculating the self force in Schwarzschild spacetime
by mode-sum regularization}

\author{Leor Barack and Amos Ori}
\address {Department of Physics,
          Technion---Israel Institute of Technology, Haifa, 32000, Israel}
\date{\today}
\maketitle


\begin{abstract}

We outline a method for calculating the self force (the "radiation reaction
force") acting on a scalar particle in a strong field orbit in a
Schwarzschild spacetime.
In this method, the contribution to the self force associated with each
multipole mode of the particle's field is calculated separately,
and the sum over modes is then evaluated, subject to a certain
regularization procedure.
We present some explicit results concerning the implementation of the
calculation scheme for a static particle, and also for a uniform circular
motion, on the Schwarzschild background.

\end{abstract}


This short manuscript deals with the calculation of the self force acting
on a small charged object (a "particle") moving in a Schwarzschild spacetime.
We shall briefly outline here a method, recently proposed by one of us (A.O.)
\cite {dublin}, and present some results.
This method involves the calculation of the contribution to the self-force
due to each individual field's multipole, and the summation
over these contributions, subject to a certain regularization procedure
(outlined below).
For simplicity we shall consider here a scalar charge. We expect,
however, that the generalization to the electromagnetic
self force will be almost straightforward. The generalization
to the Kerr case and to the gravitational self force is likely to be
possible as well.
A closely related approach to the regularization of the self force,
also based on the multipole expansion, is currently being studied by Lousto
\cite{lousto}. For an overview of further related works, the reader may
refer, e.g., to Ref.\ \cite{qw99}.

The main challenge in calculating the self force is to deal with
the various infinite quantities associated with the divergence of the
field at the particle's location.
Previously it has been proposed \cite{ori} that radiation reaction effects
on the orbit may be calculated by evaluating the contribution of each
Fourier-multipole mode $\ell mw$ of the field to the radiative evolution,
and then summing over all modes. This approach has two advantages:
First, each individual mode of the field turns out to be continuous (and the
corresponding self force to be bounded) even at the particle's location.
Secondly, calculating each $\ell mw$ mode of the field becomes a relatively
simple task, as it only requires the solution of an ODE.
In Ref. \cite{ori} this sum-over-modes approach has been proposed
for the calculation of the adiabatic,
orbit-integrated, evolution rate of the three
constants of motion in Kerr: The Energy $E$, the azimuthal angular momentum
$L$, and the Carter constant $Q$. One might have hoped that the same
method could also be used for calculating the momentary self force.
Unfortunately, at least for the self force, this naive procedure turns out,
in general, inapplicable: Although each mode yields a finite contribution,
the sum over all modes is found to diverge. This is the situation even in
the simple case of a static particle in a flat space: For such a particle,
located at a distance $r_0$ from the origin of coordinates
(with respect to which the spherical harmonics are defined), the
contribution of each multipole $\ell$ to the radial component of the self force
is the same --- $-q^2/(2r_0^2)$, where $q$ is the particle's scalar charge.
Obviously, the sum over $\ell$ diverges.\footnote {Note, however,
that the mode sum for the adiabatic, orbit-integrated, evolution rate of
$E$ and $L$ does converge
\cite{ori}. It is not clear yet whether the corresponding
mode sum for $Q$ converges or not.}
The regularization scheme outlined here is aimed to overcome this type of
divergence (though in a nontrivial black-hole spacetime).

Practically, in this approach the calculation of the self force
is composed of two separate parts:
\newline
(i) The numerical part: solving the appropriate ODE for each mode $\ell mw$
of the field, and evaluating each mode's contribution to the self
force\footnote {We use here $F_{\mu}=q\phi_{,\mu}$ as the basic expression
for the force applied on a charged particle by the scalar field.
Throughout this manuscript we shall often omit the vectorial index
of the force and related quantitied for notational brevity.}
--- a quantity we denote by $F^{\ell m\omega}$.
(Alternatively, one may numerically solve the 1+1 PDE in the time
domain, for each $\ell$ and $m$.)
\newline
(ii) The analytic part: applying a certain regularization procedure to the
mode sum. This procedure involves the analytic calculation of certain
regularization parameters, as we describe below.
\newline
This manuscript deals with the second part, namely, the analytic
regularization scheme. The first part --- the numerical determination
of $F^{\ell m\omega}$ --- is currently being implemented by Burko
for several physical scenarios \cite {burko1,burko2,burko3,Amaldi}.

The analytic regularization scheme has already been fully implemented
by one of us (A.O.)
for circular orbits in the Schwarzschild geometry \cite{unpublished}.
Barack and Ori \cite {bo} are currently developing the scheme for
noncircular orbits (focusing on radial orbits at the moment).
Since our regularization method has already been implemented
in practical calculations (for static \cite {burko1} and circular \cite {burko2}
orbits in the Schwarzschild geometry), we found it useful to provide a short
account of the regularization scheme, and to briefly describe the main
results already obtained.
A more detailed account, including a systematic development of the method
and a detailed calculation of the various parameters involved, will be
given in Ref. \cite {bo}.

As was shown by Quinn and Wald \cite{qw99} (see also \cite{mino,qw97}),
the physical self force is (in vacuum) the sum of two parts: (i) A local,
Abraham-Lorentz-Dirac (ALD)-like term, and (ii) a "tail" term
$F^{\rm (tail)}$, associated with the tail part of the Green's function.
The local term is trivial to calculate.
We shall therefore focus here on the tail term.
This term may be expressed as
\begin{equation}\label{eq1}
F^{\rm (tail)}\equiv\lim_{\epsilon\to 0^+}F_{\epsilon},
\end{equation}
where $F_{\epsilon}$ denotes the contribution to the force
(evaluated at $\tau=0$) from the part $\tau\leq -\epsilon$ of
the particle's worldline. Decomposing $F_{\epsilon}$ into $\ell$-modes,
one finds
\begin{equation}\label{eq2}
F^{\rm (tail)}= \lim_{\epsilon\to 0^+}\sum_{\ell}F^{\ell}_{\epsilon} =
\lim_{\epsilon\to 0^+}\sum_{\ell}\left(F^{\ell}-\delta F^{\ell}_{\epsilon}
\right).
\end{equation}
Here, $F^{\ell}_{\epsilon}$, $\delta F^{\ell}_{\epsilon}$,
and $F^{\ell}$ denote the force associated with the $\ell$-multipole of
the field, sourced, respectively, by the interval $\tau \leq -\epsilon$,
the interval $\tau > -\epsilon$, and the entire world-line.
The quantity $F^{\ell}$ may be identified with the sum over $m,\omega$
of the contributions from all modes $\ell,m,\omega$ for a given $\ell$.
(Recall that in calculating a field's mode $\ell,m,\omega$
one takes the source term to be the {\it entire} worldline. Since we are
using the retarded Green's function, the part $\tau > 0$ does not contribute.
However, the interval from $-\epsilon$ to $0^+$ does contribute.
Essentially, it is this part which is responsible to the instantaneous,
divergent, piece of the Green's function, which should be
removed from the expression for $F^{\rm (tail)}$.)

A clarification is required here concerning the meaning of the last
equality in Eq.\ (\ref{eq2}): Let $r_0$ denote the value of $r$ at the
self force evaluation point. Then, the quantity $F^{\ell}_{\epsilon}$ is
well defined at $r=r_0$. The situation with $F^{\ell}$ and $\delta
F^{\ell}_{\epsilon}$ is more delicate, however. Each of these quantities
has a well defined value at the limit $r\to r_0^-$, and a well defined
value at the limit $r\to r_0^+$. Generically, for the $r$-component (and
in some cases also for other components) these two one-sided values are
not the same. Eq.\ (\ref{eq2}) should thus be viewed as an equation for
either the limit $r\to r_0^-$ of the relevant quantities (i.e. $F^{\ell}$
and $\delta F^{\ell}_{\epsilon}$), or the limit $r\to r_0^+$ of these
quantities. Obviously, this equation is also valid for the {\it averaged
force}, i.e. the average of these two one-sided values. In what follows we
shall always consider the averaged force.
[Of course, the final result of
the calculation, $F^{\rm (tail)}$ (which has a well defined value at the
evaluation point), should be the same regardless of whether it is derived
from its one-sided limit $r\to r_0^-$, or from $r\to r_0^+$, or from their
average.]

Next, we seek an $\epsilon$-independent function $h^{\ell}$, such that the
series $\sum_{\ell}(F^{\ell} - h^{\ell})$ converges.
Once such a function is found, then Eq.\ (\ref{eq2}) becomes
\begin{equation}\label{eq3}
F^{\rm (tail)}=\sum_{\ell}\left(F^{\ell}-h^{\ell}\right)
-\lim_{\epsilon\to 0^+}\sum_{\ell} \left(\delta
F^{\ell}_{\epsilon}-h^{\ell}\right).
\end{equation}
In principle, $h^{\ell}$ can be found by investigating the asymptotic behavior
of $F^{\ell}$ as $\ell \to \infty$. It is also possible, however, to derive
$h^{\ell}$ from the large-$\ell$ asymptotic behavior of
$\delta F^{\ell}_{\epsilon}$ (the latter and $F^{\ell}$ must have the same
large-$\ell$ asymptotic behavior, because their difference yields a convergent
sum over $\ell$). In addition to $h^{\ell}$, the investigation
of $\delta F^{\ell}_{\epsilon}$ should also produce the parameter
\begin{equation}\label{eq4}
d \equiv \lim_{\epsilon\to 0^+}\sum_{\ell} \left(\delta
F^{\ell}_{\epsilon} - h^{\ell}\right),
\end{equation}
which is required for the calculation of $F^{\rm (tail)}$ in Eq.\
(\ref{eq3}) .

From Eqs. (\ref{eq3}) and (\ref{eq4}) it is obvious that
we only need the asymptotic behavior of $\delta F^{\ell}_{\epsilon}$
in the immediate neighborhood of $\epsilon=0$.
It is therefore possible to analyze
$\delta F^{\ell}_{\epsilon}$ using local analytic methods.
In particular, we can apply a perturbation
analysis to the $\ell$-mode field equation (in the time domain).
To that end, we express the $\ell$-mode
effective potential $V^\ell(r)$ as a small perturbation $\delta V^\ell(r)$
over the value of $V^\ell(r)$ at the evaluation point, $V^\ell_0\equiv
V^\ell(r=r_0)$. This perturbative approach is used to analyze
$G^\ell\equiv G^\ell [x^\alpha,x^\alpha_p(\tau)]$
(the $\ell$-mode Green's function),
where $x^\alpha$ is the evaluation point and $x^\alpha_p(\tau)$
describes the particle's worldline.
Expressing $G^\ell$ as a function of $\tau$ and $z\equiv \tau \ell$,
the perturbation analysis provides an expression for $G^\ell$ as a
power series in $\tau$ (with $z$-dependent coefficients). Only the
first three terms, i.e. the terms up to order $\tau^2$, are required for
the calculation of the self force
(recall that eventually we take the limit $\epsilon \to 0$; as it turns
out, in this limit the contributions from all higher-order terms vanish).
The perturbation analysis yields explicit analytic expressions for the
required three expansion coefficients (as functions of $z$), which we
present in Ref.\ \cite{bo}.

We next construct $\delta F^{\ell}_{\epsilon}$ from $G^\ell$
(essentially by integrating the latter's gradient from $\tau=-\epsilon$
to $\tau=0$). One then obtains the large-$\ell$ asymptotic form
$\delta F^{\ell}_{\epsilon}=a \ell +b +c\ell ^{-1}+O(\ell^{-2})$,
in which the parameters $a$, $b$, and $c$ are independent
of $\ell$ and $\epsilon$.
(These parameters do depend on the evaluation point,
and also on the orbit --
through the values of the particle's four-velocity and its first few
proper-time derivatives there.)
The regularization function $h^{\ell}$ thus takes the form
\begin{equation}\label{eq5}
h^{\ell}=a \ell +b +c\ell^{-1}.
\end{equation}
Consequently, the tail part of the self force is found to be
\begin{equation}\label{eq6}
F^{\rm (tail)}=\sum_{\ell}\left(F^{\ell}-a\ell-b-c\ell^{-1}\right)-d.
\end{equation}
The implementation of this regularization scheme thus amounts to analytically
calculating the four regularization parameters $a$, $b$, $c$, and $d$,
using the above mentioned perturbation expansion for the Green's function.
(Note that there exist
four such parameters for each vectorial component of the force.)

We shall now present the main results already obtained for the values of
the above regularization parameters in various cases.
These results concern the $r$-component of the self
force, and are valid for either geodesic or non-geodesic orbits.
It can be shown that the parameter $a$ vanishes for all orbits in
Schwarzschild \cite{unpublished}, and preliminary results indicate
that the same holds for $c$.
(Recall that we consider here the averaged force: the
``one-sided'' values of $a$ are, in general, nonzero \cite{bo}.)
On the other hand, the parameters $b$ and $d$ are generically nonvanishing.
In what follows we give the values of these parameters for some physical
scenarios.\footnote{We use metric signature $(-+++)$ throughout.}

{\em For static or circular orbits} the parameter $d$ (as well as $a$ and
$c$) is found to vanish \cite{unpublished,bo}:
\begin{equation}\label{eq8}
a_r=c_r=d_r=0 \quad \text{for static or circular orbits},
\end{equation}
where the sub-index $r$ indicates that these parameters are associated
with the covariant $r$-component of the self force.
Hence, in these cases one finds\footnote
{For static or circular orbits, the $r$-component of the local,
ALD-like, term of the total self force vanishes. Thus, Eq.\ (\ref{eq9})
describes, if fact, the ($r$-component of the) {\em total}
self force.}
\begin{equation}\label{eq9}
F_r^{\rm (tail)}=\sum_{\ell}\left(F_r^{\ell}-b_r\right)\quad
\text{(static or circular orbits)}.
\end{equation}
For a static particle, the parameter $b_r$ is given by \cite{bo}
\begin{equation}\label{eq10}
b_r=
-\frac{q^2}{2r_0^2}\left(\frac{r_0-M}{r_0-2M}\right)\quad
\text{(static particle)},
\end{equation}
where $M$ is the black hole's mass.
For a circular orbit, $b_r$ takes the form \cite{unpublished}
\begin{equation}\label{eq11}
b_r=-\frac{q^2}{2r_0^2}\;\frac{1}{u^t\sqrt {g_{tt}}}\; \left(
{2\,I_a-\frac{r_0-3M}{r_0-2M}\,I_b} \right)\quad \text{(circular orbit)},
\end{equation}
where
\begin{equation}\label{eq12}
I_a=F(1/2,1/2;1;v^2)\quad , \quad
I_b=F(1/2,3/2;1;v^2) ,
\end{equation}
$F$ denotes the hypergeometric function, and $v$ is the tangential
velocity with respect to a static observer:
$v=\frac{d\varphi }{dt}\,\frac{r_0}{\sqrt {g_{tt}}}$
(for an equatorial orbit).

Recently, Burko numerically calculated $F^{\ell m\omega}$ for static
\cite {burko1} and circular \cite {burko2} orbits, and used the
above regularization scheme to calculate the self force.
The above expressions (\ref{eq10},\ref{eq11})
for $b_r$ (as well as the vanishing of $a_r$ and $c_r$)
are in excellent agreement with the large-$\ell$ limit of his
numerically-deduced $F^{\ell}$ (obtained by summing $F^{\ell m\omega}$
over $m$ and $\omega$).

{\em For radial motion in flat space} (i.e., motion with fixed angular
coordinates $\theta$ and $\varphi$), the parameter
$b_r$ is found to be \cite{bo}
\begin{equation}\label{eq13}
b_{r}=-\frac{q^2}{2r_0^2}\left(1-\dot r^2+r_0 \ddot r\right)\quad
\text{(radial motion in Minkowski)},
\end{equation}
where a dot denotes differentiation with respect to proper time.
In this case, the parameter $d$ does not vanish, in general, and
one finds \cite{bo}
\begin{equation}\label{eq14}
d_{r}=\frac{1}{3}q^2 \left(\dot{a}_r-u_ra^2\right)\quad
\text{(radial motion in Minkowski)},
\end{equation}
where $u_\alpha$ and $a_\alpha$ are the four-velocity and four-acceleration,
respectively, and $a^2\equiv a^\alpha a_\alpha$.
Remarkably, $d_r$ thus equals the $r$-component of the ALD-like local term
(given in  Ref.\ \cite{qw99}).
As a consequence, this contribution from the tail term will {\em cancel out}
the $r$-component of the local ALD-like term in the expression for the total
self force.

Generalization of the above results, Eqs.\ (\ref{eq13}) and (\ref{eq14}),
to general radial motion in a Schwarzschild spacetime
(and also in a more general class of static spherically symmetric spacetimes)
will be presented in a forthcoming paper \cite{bo}.




\begin{references}

\bibitem {dublin} A. Ori, talk given in the workshop on
Black Holes and Gravitational Waves, UCD, Dublin, August 1999.

\bibitem {lousto} C. O. Lousto, {\em Pragmatic approach to gravitational
radiation reaction in binary black holes} (in preparation).

\bibitem {qw99} T. C. Quinn and R. M. Wald, Phys.\ Rev.\ D.\
{\bf 60}, 064009 (1999).

\bibitem {ori} A. Ori, Phys.\ Lett.\ A{\bf 202}, 347 (1995);
                       Phys.\ Rev.\ D{\bf 55}, 3444 (1997).

\bibitem {burko1} L. M. Burko, Class.\ Quantum Grav.\ (in press),
gr-qc/9911042.

\bibitem {burko2} L. M. Burko, {\em Self force on particle in orbit
around a black hole}  (submitted to Phys.\ Rev.\ Lett.).

\bibitem {burko3} L. M. Burko, {\em Importance of conservative
self forces for binaries undergoing radiation damping} (in preparation).

\bibitem {Amaldi} L. M. Burko, in Proceedings of the 3rd
Amaldi Conference on Gravitational Waves, edited by Sydney Meshkov
(to be published), also gr-qc/9911089.

\bibitem {bo} L. Barack and A. Ori, in preparation;
see also gr-qc/9912010.

\bibitem {unpublished} The detailed implementation of the regularization
scheme to circular and other non-radial trajectories will be published
elsewhere.

\bibitem {mino} Y. Mino, M. Sasaki, and T. Tanaka, Phys.\ Rev.\ D.\
{\bf 55}, 3457 (1997).

\bibitem {qw97} T. C. Quinn and R. M. Wald, Phys.\ Rev.\ D.\
{\bf 56}, 3381 (1997).

\end{references}
\end{document}